# A Model Study of an All-Digital, Discrete-Time and Embedded Linear Regulator

Saad Bin Nasir, Student Member, IEEE, Arijit Raychowdhury, Senior Member, IEEE

*Abstract*— With an increasing number of power-states, finer-grained power management and larger dynamic ranges of digital circuits, the integration of compact, scalable linear-regulators embedded deep within logic blocks has become important. While analog linear-regulators have traditionally been used in digital ICs, the need for digitally implementable designs that can be synthesized and embedded in digital functional units for ultra-fine grained power management has emerged. This paper presents the circuit design and control models of an all-digital, discrete-time linear regulator and explores the parametric design space for transient response time and loop stability.

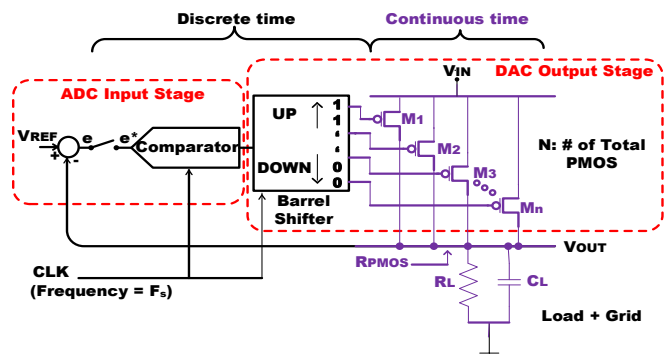

Fig. 1. Proposed all-digital discrete-time LDO with an embedded comparator, a barrel shifter and PMOS array.

## I. INTRODUCTION

WITH the growing need for higher energy efficiency in digital systems-on-chip (SoCs) and microprocessors, power management has emerged as a major design constraint. Both DC-DC conversion and voltage regulation continue to be actively researched for efficiency, compactness and faster response to provide ultra-fine grain spatio-temporal power management. On-chip power delivery networks are implemented in a hierarchical manner, combining slower and more efficient switching regulators with faster and less efficient linear regulators, to address power "hotspots" across multiple-voltage domains and wide dynamic operation [1]. Traditional low drop-out (LDO) regulators have been analog in nature and employ a high-gain error amplifier to provide regulation. They provide high bandwidth, low ripple, fast response times and high power supply rejection (PSR) [2], [3]. However, the use of analog design principles do not allow operation at low input and control voltages and are difficult to integrate as collaterals embedded deep within a digital functional unit. This has inspired the design of digital implementations of LDOs [4]-[7] targeted for digital load circuits. The rationale behind such designs is to convert the control section of an LDO into a digitally implementable circuit which is easier to integrate in scaled nodes. Secondly, it allows the designer to replicate and distribute such regulators in larger numbers on the die to provide ultra-fine grained spatio-temporal power management.

Digitally implementable LDOs are implemented using either (i) continuous-time [4], [6] or (ii) discrete-time control. Discrete-time control [5], [7] uses a master clock that synchronizes all the data movement in the control portion of the regulator. This paper builds on top of [7] and provides an in-depth analysis of the discrete-time all-digital LDO, taking into account both the transient and steady state operations. The key contributions of this brief are:

1. Development of a z-domain control model of the LDO illustrating the relationship between key design parameters and transient response.
2. Development of a steady state model for assessment of limit cycle oscillations in the LDO.
3. Design space exploration of the LDO for performance, power efficiency and stability.

## II. DESIGN OF DISCRETE-TIME ALL-DIGITAL LDO

Before presenting the model study of the digital LDO, let us explore the key design elements that have been proposed in [7] (Fig. 1). Design and simulations were performed using the IBM 130nm LP process. The regulated (or output) voltage ($V_{OUT}$) is synchronously sampled by a single-bit comparator which provides a binary ($V_{REF}>V_{OUT}$ or $V_{OUT}>V_{REF}$) signal. The comparator output (which in essence represents a 1-bit ADC) controls a programmable barrel shifter with parallel outputs to switch ON (or OFF) the power MOSFETs.

### A. Comparator and shift stages

As proposed in [7], a single bit comparator provides a compact and efficient design. A clocked sense amplifier based topology is employed [8]. It obviates the need for constant bias current

Manuscript submitted on XX. This work was funded by the Semiconductor Research Corporation (Task no 1836.140), gifts from Qualcomm Inc. and Intel Corp. and the Fulbright Fellowship Program.
The authors are with the School of Electrical and Computer Engineering at Georgia Institute of Technology, Atlanta, USA.

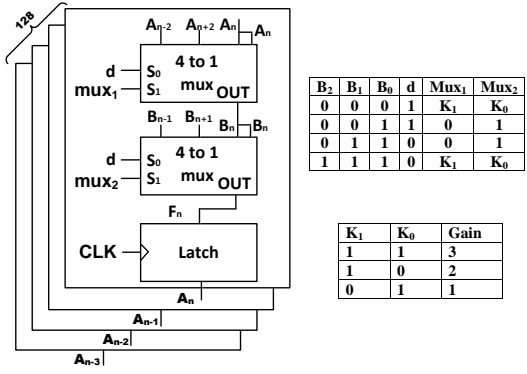

Fig. 2. Design schematic of a 128 bit barrel shifter using 4x1 MUXes and latches to provide programmable magnitude and direction of shift.

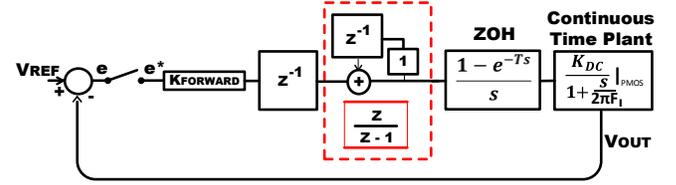

Fig. 3. Hybrid control model illustrating the transient dynamics of the proposed LDO.

of a clock-less comparator. During the negative phase of the clock, the output is pre-charged to $V_{DD}$. At the positive clock edge, the faster discharging node resets the discharging process of the other one and the comparator output results in 1 if $V_{IN}<V_{REF}$ or 0 if $V_{IN}>V_{REF}$. The decision is latched using an SR latch. A bi-directional barrel shifter takes input from the comparator and turns on/off a fixed number of power PMOS devices using a thermometer code through a barrel shifter. It is set using a register programmable gain. In the present implementation, the shift register is 128 bits wide. A programmable gain of {+3,+2,+1,-1,-2,-3} is implemented through two levels of multiplexing. At the $n^{th}$ barrel shifter output, the first level MUX takes $A_n$, $A_{n-2}$ and $A_{n+2}$ as input to produce one of the intermediate values ($B_n$) for the next level of MUXes. This allows a shift of {0, +2, or -2}. The other inputs on the second MUX levels are $B_{n-1}$ and $B_{n+1}$ which allow an additional {+1, -1, or 0} shifts, thus realizing the whole range of programmable gain as depicted in Fig. 2.

*B. Power PMOS Array*

A total of 128 equally sized PMOS devices are connected as pass devices to regulate the output voltage. The exact size of each PMOS is dictated by the underlying process technology, the target voltage range and the load current specifications. In the current design, the PMOS array has a total area of 51.2 μm and is capable of delivering a maximum of 3.5 mA at a nominal output voltage of 0.7 V from a supply voltage of 1 V. The resistance offered by digital load circuit ($R_L$) and the on-chip capacitance at the output from the grid, including all the parasitic and on-chip decoupling capacitances ($C_L$), account for the total load. . A grid and load capacitance of 1nF [6] has been assumed. The design can be fully integrated without the need for external capacitors

III. HYBRID CONTROL MODEL FOR THE REGULATOR

Small signal dynamics are modeled in the *z*-domain to account for the all-digital control. The synchronous comparator can be represented by a subtraction followed by an impulse voltage sampler running at the clock frequency ($F_s$). It gives a unit-less discrete time error sample at time nT, as:

$$e^* = V_{REF}(nT) - V_{OUT}(nT) \quad (1)$$

where $T=1/F_s$. At every clock cycle the error is either +1 or -1 and in steady state it represents a bang-bang control. The barrel shifter acts as a perfect discrete time integrator as shown in Fig. 3. The output of the shifter, which is a thermometer coded digital word, $D(nT)$, represents the number of PMOS devices that are on at any time instant, nT. This discrete integration in the form of a difference equation is:

$$D(nT) = D((n-1)T) + K_{FORWARD}\, e((n-1)T) \quad (2)$$

The digital word is applied after a single cycle clock delay following synchronous operation of the LDO. $K_{FORWARD}$ is the forward gain and is adjusted by setting the step-size in the barrel shifter {+3,+2,+1,0,-1,-2,-3}. From (2) we obtain:

$$D(z)=K_{FORWARD}\frac{1}{z-1}e(z) \quad (3)$$

The output of the shifter interfaces with a continuous time plant (load circuit) through a zero order hold (ZOH). Direct transformation is employed to convert s-domain transfer function to z-domain giving (4).

$$P(z) = \frac{K_{OUT}(1-e^{-F_l T})}{F_l}\left(\frac{1}{z-e^{-F_l T}}\right) \quad (4)$$

The load pole is at $F_l = \frac{1}{(R_L||R_{PMOS})C_L}$ and $K_{OUT}$ is the gain in the plant transfer function. $K_{OUT}$ is proportional to $I_{PMOS}$ and can be expressed as:

$$K_{OUT}= K_{DC}*I_{pmos} \quad (5)$$

Where $I_{pmos}$ is the current of a single 'ON' PMOS and is equal for equally sized PMOS devices in the array and $K_{DC}$ is a DC proportionality constant. Thus using (3), (4) and (5) the open loop forward path transfer function of the digital LDO is written in the *z*-domain as

$$G(z) = \frac{K_{FORWARD}\, K_{DC}\, I_{PMOS}(1-e^{-F_l T})}{F_l}\frac{1}{(z-1)(z-e^{-F_l T})} \quad (6)$$

This represents a second order system. The pole at (1,0) on the unit circle comes from the digital integration (equivalent to a single tap IIR filter) whereas, the second pole at $e^{-F_l T}$ is function of both load ($F_l$) and sampling frequency ($F_s$). Using unity feedback, the overall closed loop transfer function of the digital LDO in *z*-domain is:

$$H(z)=K\frac{1}{z^2-(1+e^{-F_l T})z+K(1-e^{-F_l T})+e^{-F_l T}} \quad (7)$$

where the open loop gain $K = \frac{K_{FORWARD}\, K_{DC}\, I_{PMOS}(1-e^{-F_l T})}{F_l}$. Root locus of the closed loop poles as the open loop gain (K) increases as shown in Fig. 4a for two different load conditions. Similar to second order continuous time systems, the current system may also become oscillatory and can exhibit instability for high values of K.

*A. Comparison of transient model and simulation*

In this sub-section, we establish the notion of stability in digital LDOs and specify different parameters which affect its transient performance. Obtaining phase margin (PM) through Bode plots is not feasible in perfect integration systems like in a digital LDO [9]. Therefore, PM is approximated in time domain. Given the percentage overshoot ($O_s$), the damping factor ($\zeta$) of the system is calculated using:

$$O_s = e^{-\pi\zeta/\sqrt{1+\zeta^2}} \quad (8)$$

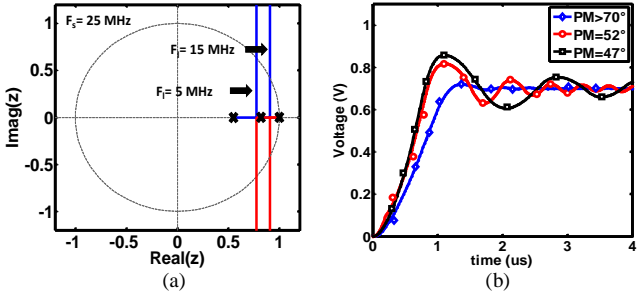

Fig. 4. (a) Root locus for increasing open loop gain (K) with constant $F_s$ and $F_l$=5, 15 MHz obtained through the model. Load conditions change the open loop poles and the breakaway point on the root locus plot. (b) Phase margin (PM) decreases with increase in overshoot ($O_s$) as obtained through HSPICE simulations for decreasing $F_l$ with constant load conditions and $F_s$.

Then PM is given as $\zeta*100$. This approximation works well for second order systems to check for any potential instability at the boundary specifications. Fig. 4b shows changing PM for different load conditions of LDO obtained through HSPICE simulations. Increasing the open loop gain (K) moves the closed loop poles on the locus decreasing PM as shown in the Fig. 4. This establishes an upper bound on K which can be ascertained through the model for a given $F_s$ and $F_l$. The two control knobs which a designer can exercise to change K are the barrel shifter gain, $K_{FORWARD}$ and the $I_{PMOS}$. The size of each PMOS in the array is dictated by the load current specification during design time. During run-time $K_{FORWARD}$ can be used as a control parameter for variable proportional gain in the forward path. Fig. 5 illustrates the movement of the system poles and corresponding performance improvement as $K_{FORWARD}$ is increased. Change of $F_s$ brings the open loop output pole closer to the integrator pole, thus decreasing the PM. Hence, even with a low value of K, the designer can obtain faster transient response at an increased $F_s$. Step response for similar load conditions and $K_{FORWARD}$ but increased $F_s$ are compared in Fig. 6a. A 5X increase in $F_s$ is simulated for a load step of 1.5 mA from 0.9 to 2.4 mA. It shows over a 6X decrease in the rise time, from 0 V to within 5% of 700 mV, of regulated output voltage. Correspondingly, we see a change in the position of open loop poles for increasing values of $F_s$ as verified by the z-plane pole-zero plots in Fig. 6(b-d).

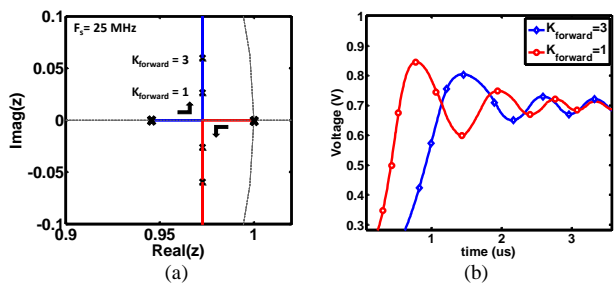

Fig. 5. (a) Shift in closed loop pole locations for $K_{FORWARD}$ = 1, 3 obtained through proposed z-domain control model (b) Equivalent transient response for variation of barrel shifter gain ($K_{FORWARD}$) using HSPICE simulations with $F_s$ = 50MHz, $I_{PMOS}$ = 1.5mA and $C_L$ = 1nF.

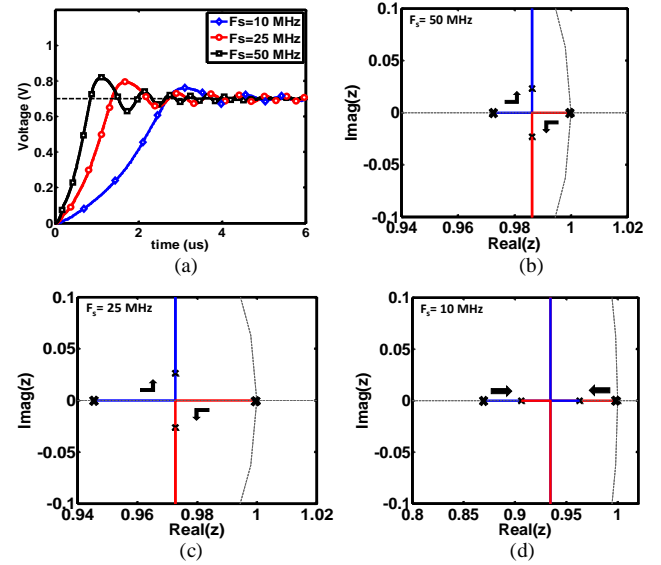

Fig. 6. (a) Transient Response for variation of sampling frequency ($F_s$) using HSPICE. (b) Equivalent Shift in closed loop pole locations for $F_s$ = 50, 25, 10 MHz is shown in (b)-(d) with $K_{FORWARD}$ = 1, $I_{PMOS}$ = 1.5mA and $C_L$ = 1nF.

From HSPICE simulations we note that a 3X change in $K_{FORWARD}$ can achieve the same improvement in the transient response as a 2.5X increase in $F_s$ (comparing Figs. 5 and 6). While $K_{FORWARD}$ determines the position of the closed loop poles on the root locus, $F_s$ changes the position of one of the open loop poles and can thus actuate a faster transient response.

## IV. STEADY STATE REGULATOR MODEL

Although the linear small signal models, as described in Section III allow us to qualitatively and quantitatively analyze the transient response and stability criteria in digital LDOs, such models fail to comprehend the steady-state behavior, in particularly the role of limit cycle oscillations. Limit cycle oscillations are observed in digital LDOs [5] because of the hard quantization at the comparator input. Further, the delay between the sampling instant and the time of actuation through the PMOS devices plays a critical role in this regard.

A nonlinear sampled feedback system is proposed to capture this behavior (Fig. 7). The comparator in the feedback loop exhibits the transfer characteristics of an ideal relay with zero dead-time. All other components take the same form as presented in Section III. Such a system with a hard nonlinearity i.e., an ideal relay, shows limit cycle oscillations and lends itself to be analyzed in terms of describing functions [10]. At every positive clock edge, the comparator (or relay) will turn on or off a PMOS giving a periodic output. For example, a sequence {+1 +1 +1 -1 -1 -1} exhibits a limit cycle and will be referred to as a mode-3 oscillation. Here mode refers to the number of switching power PMOSes in steady state. Any other explicit delays in the loop can also be incorporated in the following holistic analyses.

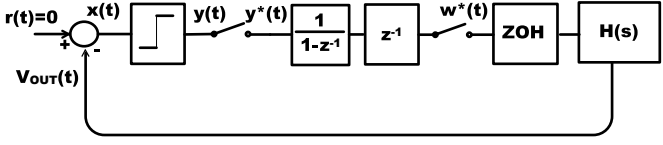

Fig. 7. Steady state LDO model to analyze time-domain waveforms at different points in the control loop.

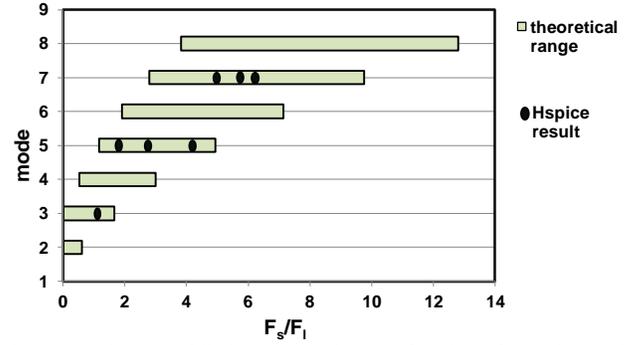

Fig. 8. The existence of limit cycle modes as a function of $F_s/F_l$ in steady state of the baseline design with HSPICE results superimposed.

Following Fig. 7, let us determine the condition for mode-n to exist. In a limit cycle, the input to the relay, x(t), is sinusoidal. At y(t), 'n' PMOS devices switch in a single cycle of frequency $\omega_s/2n$ and satisfies mode-n. Each sample at y(t) is separated by a maximum of $(360/2n)°$ for mode-n. If one of the sampling instants is assumed to be at the origin then the following is the necessary criteria for mode-n to exist.

$$x(t) = A\sin\left(\frac{\omega_s}{2n}t + \varphi\right); \qquad 0 < \varphi < \frac{180°}{n} \qquad (9)$$

where A is the amplitude of the ripple and $\varphi$ is the phase term with respect to an assumed origin of x(t). y*(t) samples the relay output, y(t). The fundamental component of these samples is advanced by $(180/2n)°$. If each sample $y^*(t)$ is of amplitude D then $|y^*(t)|$ is:

$$|y^*(t)| = \frac{2D}{nT}\int_{0^-}^{nT^-} y^*(t)\sin\left(\frac{\omega_s}{2n}t + \frac{180°}{2n}\right)dt \qquad (10)$$

This integration is solved as a summation of impulses in one cycle. It should only include one of the two endpoint samples. From [10], the sampled describing function is given as:

$$N(A,\varphi) = \frac{\text{phasor representation of fundamental component of } y^*(t)}{\text{phasor representation of x(t)}} \qquad (11)$$

Then for mode-n it reduces to:

$$N(A,\varphi) = \frac{\frac{2D}{nT}\int_{0^-}^{nT^-} y^*(t)\sin\left(\frac{\omega_s}{2n}t + \frac{180°}{2n}\right)dt}{A\sin\left(\frac{\omega_s}{2n}t + \varphi\right)} \qquad 0 < \varphi < \frac{180°}{n} \qquad (12)$$

The sinusoidal response function for the linear part is given by

$$L(j\omega) = H(j\omega)Z(j\omega)S(j\omega) \qquad (13)$$

Here $S$ gives the discrete integration and the clock cycle delay. $Z$ represents the ZOH followed by $H$ as the load frequency response. $L(j\omega)$ is evaluated at the oscillation frequency of $\omega = \omega_s/2n$ which simplifies to:

$$L\left(j\frac{\omega_s}{2n}\right) = \frac{(1-e^{-j\left(\frac{\omega_s}{2n}\right)T})(e^{-j\left(\frac{\omega_s}{2n}\right)T})\angle-\tan^{-1}\left(\frac{\omega_s\tau}{2n}\right)}{(1-e^{-j\left(\frac{\omega_s}{2n}\right)T})\,j\left(\frac{\omega_s}{2n}\right)\sqrt{1+\left(\frac{\omega_s\tau}{2n}\right)^2}} \qquad (14)$$

For a limit cycle to exist, according to the Nyquist criterion

$$N(A,\varphi)L\left(j\frac{\omega_s}{2n}\right) = -1 = 1\angle-180° \qquad 0 < \varphi < \frac{180°}{n} \qquad (15)$$

Comparing the phase terms in (15) and applying the inequality gives the bounds of $\frac{F_s}{F_l}$ for which mode-n can exist.

Similarly, all the necessary bounds of relevant modes can be calculated for the proposed LDO by evaluating (15) within the phase constraints. Loop gain also needs to be greater than 0 dB to achieve the sufficient conditions for a given mode to exist. Loop gain is numerically computed through simulations and model calibration. We note a good corroboration of the proposed model with simulations. Control propagation delay in the loop and any delay in the output voltage grid are negligible compared to the clock frequency (typically 10-100 MHz). Similar necessary conditions for different modes can be found following the same analysis. Fig. 8 summarizes the results for the given design.

## V. Parametric Design Trade-offs And Optimizations

From the control model described in section III and the steady state model described in section IV, it becomes clear that the choice of $F_s$ plays a major role in ascertaining both the stability and performance of a digital LDO. Further, as $F_S$ increases the mode of limit-cycle oscillation also increases.

The delay of one clock cycle in the baseline design (see Fig. 7) can be reduced by changing the clocking mechanism. If the comparator samples on the positive clock edge and the barrel shifter updates its output at the following negative edge then the transport delay reduces to half of a clock cycle. Following a procedure similar to that of Section IV, we can determine the bounds on the limit cycle modes for a given $F_S/F_l$. We note a reduction of the limit cycle mode at iso-$F_S/F_l$. The results from the proposed model and simulations are summarized in Fig. 9.

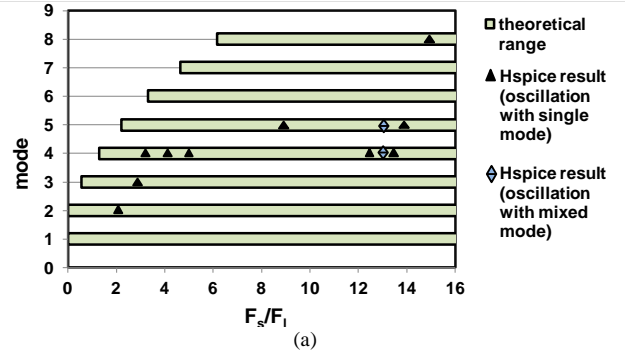

(a)

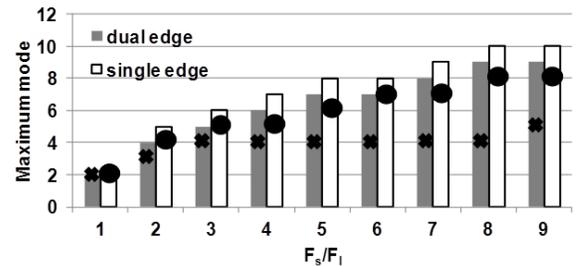

(b)

Fig. 9. (a) The theoretical bounds of limit cycle modes as a function of $F_s/F_l$ for dual edge LDO in steady state with simulation results superimposed. Mixed-mode refers to limit cycles exhibiting multiple modes existing simultaneously in the steady state. (b) Decrease in the maximum mode for $F_s/F_l$ of dual edge LDO compared to the baseline version (HSPICE results).

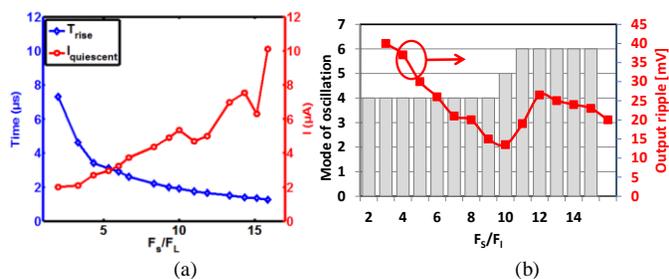

Fig. 10.  (a) Simulation results for rise time ($T_r$) and quiescent current ($I_q$) with increasing $F_s/F_l$. $T_r$ is calculated in response to a step reference change from 0 to 700 mV in 50 ps and $I_q$ is the current drawn by control circuitry for $I_{load}$ = 1mA, $C_L$ = 1nF.  (b) Simulated change in output ripple with increasing $F_s$ relative to a fixed $F_l$ with observed modes ($I_{load}$ = 1mA, $C_L$ = 1nF).

A design space exploration is carried out for optimally designing a practical digital LDO. An exponential reduction in the rise time ($T_r$) is achieved for higher values of $F_s/F_l$ with diminishing returns as $F_s/F_l$ approaches 20. Current efficiency decreases linearly with increasing $F_s$ but drops exponentially as $F_s/F_l$ ratio approaches 20. HSPICE simulation results for dual edge LDO are summarized in Fig. 10a, where the current consumed by the LDO's control logic increases from 2µA to 10µA as $F_s/F_l$ increases from 2 to 17. In the same span, $T_r$ improves by more than 600%. An optimal trade-off between power consumption and transient performance is obtained when $F_s/F_l$ ranges from values of 5 to 10. On the other hand, $F_s$ has a non-monotonic effect on the output voltage ripple. Increasing $F_s$ tends to reduce the output ripple as the load capacitor performs better filtering of the steady state oscillations. However, it increases the mode of limit cycle oscillations. Thus the net output ripple is a resultant of these two effects. For lower values of $F_s/F_l$, an exponential decrease in the output voltage ripple is first obtained and it tends to flatten out as $F_s/F_l$ becomes higher. A large increase in the ripple is obtained when the mode increases. Fig. 10b illustrates the simulated output ripple as a function of $F_s/F_l$ for a particular design instance showing a non-monotonic behavior. Thus we note that for an optimal design, it is necessary to select $F_s$ to match $F_l$ for some specification of the output ripple. As a rule of thumb, $F_s/F_l$ in the range of 5 to 10 provides maximum performance gain while maintaining output ripple and power efficiency.

In summary, we note that the optimal trade-offs of transient rise time, power consumption and output voltage ripple is obtained when $F_s$ is 5 to 10 times higher than the value of $F_l$. Therefore, $F_s$ should track $F_l$ for optimal performance. As the output load current (and hence $F_l$) changes, a worst case or an adaptive design [7] principle can be employed. In light load conditions, a low $F_s$ can optimize power consumption and still provide regulation. The upper bound on $F_s$ is determined by the transient response time and conditions for stability. The design space ($K_{forward}$-$F_s$) shown in Fig. 11 illustrates the regions of stable operation and the iso-$T_r$ contours.

VI. CONCLUSION

An all-digital discrete time LDO for fine-grained power management of digital circuits is described in this paper. Theoretical models for analyzing the transient and steady state response of the LDO are corroborated with HSPICE simulations using a commercial process design kit.  The role of the design topologies and  parameters like clock frequency, dual clock edge triggering, barrel shifter based open-loop variable gain have been explored and shown to provide the designer with control over the transient and steady-state performance. A sampling clock frequency which is 5X-10X of the load pole frequency has been shown to provide an optimal trade-off among performance, power and steady-state output ripple.

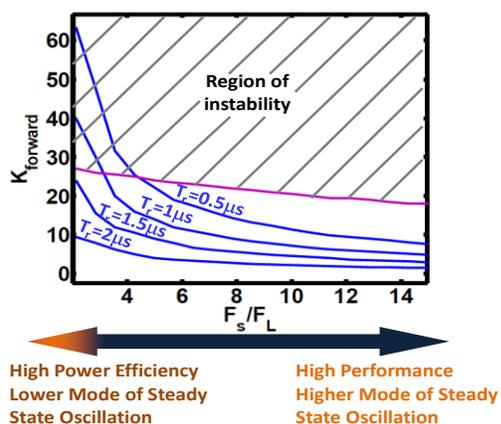

Fig. 11.   Design trade-offs in a discrete-time digital LDO on a $K_{forward}$ vs $F_s/F_l$ plot with iso-rise  time (0 to 700 mV) contours.